\documentclass[aip,pop,12pt,preprint,amsmath,amssymb,groupedaddress]{revtex4-1} 

\usepackage{bm}
\usepackage{graphicx}

\usepackage{mathptmx} 

\DeclareSymbolFont{newfont}{OML}{cmm}{m}{it}
\DeclareMathSymbol{\Varrho}{3}{newfont}{37}

\newcommand{\wt}{\widetilde}

\newcommand{\ave}[1]{{\left<#1\right>}}

\newcommand{\abs}[1]{{\left|#1\right|}}

\newcommand{\sech}{\ensuremath{\text{sech}}}
\newcommand{\csch}{\ensuremath{\text{csch}}}

\newcommand{\rmd}{\text{d}}

\newcommand{\taud}{\ensuremath{\tau_\text{d}}}

\newcommand{\tauw}{\ensuremath{\tau_\text{w}}}

\newcommand{\Phiave}{\ensuremath{\ave{\Phi}}}
\newcommand{\Phirms}{\ensuremath{\Phi}_\text{rms}}

\newcommand{\Phitilde}{\ensuremath{\widetilde{\Phi}}}

\newcommand{\aveA}{\ensuremath{\ave{A}}}
\newcommand{\Aave}{\ensuremath{\langle{A}\rangle}}

\newcommand{\Kave}{\ensuremath{\langle{K}\rangle}}

\newcommand{\Eqref}[1]{Eq.~\eqref{#1}}
\newcommand{\Eqsref}[1]{Eqs.~\eqref{#1}}
\newcommand{\Figref}[1]{Fig.~\ref{#1}}

\newcommand{\PPCF}{\textit{Plasma Phys.\ Contr.\ Fusion}}
\newcommand{\PFR}{\textit{Plasma Fusion Res.}}
\newcommand{\PHP}{\textit{Phys.\ Plasmas}}
\newcommand{\PLA}{\textit{Phys.\ Lett.~A}}
\newcommand{\PP}{\textit{Phys.\ Plasmas}}

\newcommand{\PRA}{\textit{Phys.\ Rev.~A}}
\newcommand{\PRE}{\textit{Phys.\ Rev.~E}}
\newcommand{\PS}{\textit{Phys.\ Scripta}}

\newcommand{\BTSJ}{\textit{Bell Sys.\ Tech. J.}}

\newcommand{\PRL}{\textit{Phys.~Rev.\ Lett.}}

\begin{document}

\title{Skewed Lorentzian pulses and exponential frequency power spectra}

\author{O.~E.~Garcia}
\email{odd.erik.garcia@uit.no}
\author{A.~Theodorsen}
\email{audun.theodorsen@uit.no}

\affiliation{Department of Physics and Technology, UiT The Arctic University of Norway, N-9037 Troms{\o}, Norway}

\date{\today}

\begin{abstract}
Frequency power spectra due to a super-position of uncorrelated Lorentzian pulses with a random distribution of amplitudes are considered. For pulses with constant duration, there is an exponential frequency spectrum which is independent of the degree of pulse overlap and the pulse amplitude distribution. The spectrum is furthermore shown to be unaffected by skewness of the Lorentzian pulses and even a random distribution of the pulse asymmetry parameter and its correlation with the pulse amplitude. This stochastic model provides new insight to the ubiquitous exponential spectra in  fluids and magnetized plasmas exhibiting deterministic chaos, where non-linear advection processes lead to amplitude dependent steepening of smooth pulses.
\end{abstract}

\maketitle

An intrinsic property of deterministic chaos is an exponential frequency power spectral density for the fluctuations.\cite{frish,farmer,greenside,sigeti1,sigeti2,ohtomo,mm-pre,mm-prl} This has been observed in numerous experiments and model simulations of fluids and magnetized plasmas.\cite{mm-prl,atten,libchaber,brandstater,streett,mensour,paul,mckee,hornung,pace-prl,pace-php,mm-ppcf,mrm-ppcf,zhu,decristoforo} Recently, the exponential spectrum has been attributed to the presence of uncorrelated Lorentzian pulses in the temporal dynamics.\cite{ohtomo,mm-prl,mm-pre,pace-prl,pace-php,hornung,mm-ppcf,mrm-ppcf,zhu,decristoforo} However, far from the threshold for linear instability in continuum systems, it is expected that non-linear advection processes and chaos will lead to steepening effects and randomness of the pulse parameters.\cite{bian,garcia-blob,garcia-blob2} This motivates investigations of the effect of pulse skewness and a distribution of pulse amplitudes and asymmetry.\cite{mm-ppcf,gt-l}

In this work, a stochastic model for intermittent fluctuations in physical systems based on a super-position of uncorrelated pulses is considered. By analogy with the characteristic function for stable distributions, the symmetric Lorentzian pulse is generalized to have finite skewness. The mean, variance, auto-correlation function and frequency power spectral density are all shown to be independent of a random distribution of the pulse asymmetry parameter and its correlation with the pulse amplitudes. The model presented here provides a novel framework for describing intermittent fluctuations and exponential frequency power spectral densities in fluids and magnetized plasmas.

Consider a stochastic process given by the super-position of $K$ uncorrelated pulses with a fixed shape and duration $\taud$ in a time interval of duration $T$,\cite{garcia-prl,garcia-php,garcia-psd,gt-l,rice1,parzen,pecseli,kube-php,theodorsen-php,theodorsen-ps}
\begin{equation}\label{shotnoise}
\Phi_K(t) = \sum_{k=1}^{K(T)} A_k\phi\left( \frac{t-t_k}{\taud} \right) .
\end{equation}
Each pulse labeled $k$ is characterized by an amplitude $A_k$ and arrival time $t_k$, both assumed to be uncorrelated and each of them independent and identically distributed. The number of pulses $K$ in an interval of duration $T$ is given by the Poisson distribution,
\begin{equation} \label{poisson}
P_K(K|T) = \frac{1}{K!}\left(\frac{T}{\tauw}\right)^K\exp\left(-\frac{T}{\tauw} \right) ,
\end{equation}
with mean value of the number of pulses given by
\begin{equation}
\Kave = \sum_{K=0}^{\infty} KP_K(K|T) = \frac{T}{\tauw} .
\end{equation}
Here and in the following, angular brackets denote the average of the argument over all random variables unless otherwise explicitly stated. From the Poisson distribution it follows that the waiting times between the pulses are exponentially distributed with mean value $\tauw$ and that the pulse arrival times are uniformly distributed on the time interval under consideration, that is, their probability density function is given by $1/T$.

The pulse shape $\phi(\theta)$ is taken to be the same for all events in \Eqref{shotnoise}. This function is normalized such that
\begin{equation}\label{duration}
\int_{-\infty}^{\infty} \rmd\theta\,\abs{\phi(\theta)} = 1 .
\end{equation}
Thus, each pulse contributes equally to the mean value of $\Phi_K(t)$. The integral of the $n$'th power of the pulse shape will appear frequently in the following and is defined as
\begin{equation}\label{pulseint}
I_n = \int_{-\infty}^{\infty} \rmd\theta\,\left[ \phi(\theta) \right]^n ,
\end{equation}
for positive integers $n$. It is assumed that $T$ is large compared with the range of values of $t$ for which $\phi(t/\taud)$ is appreciably different from zero, thus allowing to neglect end effects in a given realization of the process.

The normalized auto-correlation function of the pulse function is defined by\cite{gt-l,garcia-psd}
\begin{equation}\label{rhophi}
\rho_\phi(\theta) = \frac{1}{I_2}\int_{-\infty}^{\infty} \rmd\chi\,\phi(\chi)\phi(\chi+\theta) ,
\end{equation}
and the Fourier transform of this gives the frequency spectrum,
\begin{equation}\label{varrhophi}
\Varrho_\phi(\vartheta) = \int_{-\infty}^\infty \rmd\theta\,\rho_\phi(\theta)\exp(i\theta\vartheta) = \frac{1}{I_2}\,\abs{\varphi(\vartheta)}^2 ,
\end{equation}
where the transform of the pulse function is defined by
\begin{equation}\label{varphi}
\varphi(\vartheta) = \int_{-\infty}^\infty \rmd\theta\,\phi(\theta)\exp(i\theta\vartheta) .
\end{equation}
In this contribution, the auto-correlation function and the frequency power spectral density for the process defined by \Eqref{shotnoise} will be investigated for the case of skewed Lorentzian pulses.

The degree of pulse overlap, and therefore the intermittency of the process, is given by the ratio of the pulse duration and waiting times. This ratio is referred to as the intermittency parameter and defined by $\gamma=\taud/\tauw$. The mean value of the process is given by $\Phiave=\gamma I_1\Aave$ and the variance by $\Phirms^2=\gamma I_2\langle{A^2}\rangle$. In the following, the rescaled variable defined by
\begin{equation}\label{Phitilde}
\Phitilde = \frac{\Phi-\Phiave}{\Phirms} ,
\end{equation}
with vanishing mean and unit standard deviation will be considered. The auto-correlation function for a time lag $r$ is defined by
\begin{equation}
R_\Phi(r) = \langle{\Phi_K(t)\Phi_K(t+r)}\rangle .
\end{equation}
Substituting the variable defined by \Eqref{shotnoise}, the expressions for the auto-correlation function and the frequency power spectral density for the rescaled random variable defined by \Eqref{Phitilde} are given by\cite{gt-l,garcia-psd}
\begin{subequations} \label{RS_Phi}
\begin{align}
R_{\wt{\Phi}}(r) & = \rho_\phi(r/\taud) ,
\\
\Omega_{\wt{\Phi}}(\omega) & = \taud\Varrho_\phi(\omega\taud) ,
\end{align}
\end{subequations}
that is, they are simply determined by the shape and duration of the pulse function $\phi$. It is emphasized that the auto-correlation function and frequency power spectrum are independent of the amplitude distribution $P_A$ and the intermittency parameter $\gamma$, that is, the degree of pulse overlap.

The pulse shape to be considered in the following is the normalized Lorentzian function,
\begin{equation} \label{lorentzian}
\phi(\theta) = \frac{1}{\pi}\frac{1}{1+\theta^2} ,
\end{equation}
which satisfies the requirement given by \Eqref{duration}. The integral of the $n$-th power of the Lorentzian pulse shape is given by
\begin{equation} \label{InL}
I_n = \frac{1}{\pi^{n-1/2}}\frac{\Gamma(n-1/2)}{\Gamma(n)} ,
\end{equation}
where $\Gamma$ is the Gamma function. The lowest order pulse function integrals are given by $I_1=1$, $I_2=1/2\pi$, $I_3=3/8\pi^2$ and $I_4=5/16\pi^3$. From this it follows that the mean value of the stationary process is given by $\Phiave=\gamma\aveA$, the variance is $\Phirms^2=\gamma\langle{A^2}\rangle/2\pi$, and the skewness and flatness moments are\cite{gt-l}
\begin{subequations} \label{SandF-L}
\begin{align}
S_\Phi & = \frac{3}{4} \left( \frac{2}{\pi\gamma} \right)^{1/2} \frac{\langle{A^3}\rangle}{\langle{A^2}\rangle^{3/2}} ,
\\
F_\Phi & = 3 + \frac{5}{4\pi\gamma}\frac{\langle{A^4}\rangle}{\langle{A^2}\rangle^2} ,
\end{align}
\end{subequations}
clearly revealing the strong intermittency in the case of weak pulse overlap.

With the Lorentzian pulse, the normalized auto-correlation function and its Fourier transform are
\begin{subequations} \label{RvarphiL}
\begin{align}
\rho_\phi(\theta) & = \frac{4}{4+\theta^2} ,
\\
\Varrho_\phi(\vartheta) & = 2\pi\exp(-2\abs{\vartheta}) .
\end{align}
\end{subequations}
This gives the auto-correlation function and the power spectral density for the stochastic process by use of \Eqsref{RS_Phi}. It follows that the auto-correlation function is itself a Lorentzian function and therefore has algebraic tails while the power spectral density has an exponential dependence on the frequency,\cite{gt-l}
\begin{subequations}
\begin{align}\label{RPhiL}
R_{\wt{\Phi}}(r) & = \frac{4}{4+(r/\taud)^2} ,
\\
\label{SPhiexp}
\Omega_{\wt{\Phi}}(\omega) & = 2\pi\taud \exp\left( - 2\taud\abs{\omega} \right) .
\end{align}
\end{subequations}
This establishes the familiar exponential spectrum for a super-position of uncorrelated Lorentzian pulses.\cite{ohtomo,mm-pre,hornung,mm-prl,mrm-ppcf,pace-php,pace-prl,mm-ppcf,zhu,decristoforo}

The Lorentzian pulse shape can be generalized to have finite skewness based on an analogy with stable probability density functions.\cite{mm-ppcf} Since closed analytical forms are in general not known, it is defined via the inverse Fourier transform, or the characteristic function in the case of stable distributions,
\begin{equation}\label{phiinvF}
\phi(\theta;\alpha,\sigma,\lambda) = \frac{1}{2\pi}\int_{-\infty}^{\infty} \rmd\vartheta\,\varphi(\vartheta;\alpha,\sigma,\lambda)\exp(-i\theta\vartheta) ,
\end{equation}
where the Fourier transform of the pulse function is defined by
\begin{equation} \label{stabled}
\varphi(\vartheta;\alpha,\sigma,\lambda) = \exp\left( i\lambda\vartheta - \abs{\vartheta}^\alpha[ 1 - i\sigma\,\text{sign}(\vartheta) G(\vartheta,\alpha) ] \right) .
\end{equation}
The phase function is given by
\begin{equation}
G(\vartheta,\alpha) =
\begin{cases}
\tan\left(\displaystyle\frac{\pi\alpha}{2}\right) , & \text{if }\alpha\neq1 ,
\\
- \displaystyle\frac{2}{\pi}\,\ln\abs{\vartheta} , & \text{if }\alpha=1 .
\end{cases}
\end{equation}
Here $\lambda$ is a location parameter, $\sigma$ is an asymmetry parameter and $\alpha$ is referred to as the stability parameter in the context of stable distributions. For stable distributions, the stability parameter is restricted to the range $0<\alpha\leq2$ and the asymmetry parameter is restricted to the range $-1\leq\sigma\leq1$. The symmetric Lorentzian pulse shape is obtained for $\alpha=1$ and $\sigma=0$. The Fourier transform of a skewed Lorentzian pulse function is thus given by
\begin{equation}\label{FsL}
\varphi(\vartheta;1,\sigma,\lambda) = \exp(-\abs{\vartheta})\exp\left( i\lambda\vartheta-i\frac{2\sigma}{\pi}\,\vartheta\ln\abs{\vartheta} \right) .
\end{equation}
Here it is to be noted that the asymmetry and location parameters affect only the phase of this complex function. As will be shown below, this is the reason why the frequency spectrum is unaffected by the skewness of the Lorentzian pulses.

The integral of the $n$'th power of the skewed Lorentzian pulse function in general depends on both the asymmetry and location parameters,
\begin{align}
I_n(\sigma,\lambda) & = \int_{-\infty}^{\infty} \rmd\theta\,[\phi(\theta;1,\sigma,\lambda)]^n \notag
\\
& = \int_{-\infty}^{\infty} \rmd\theta\,\mathcal{F}^{-1}\left[ \mathcal{F}[\phi^n](\vartheta) \right](\theta) \notag
\\
& = \int_{-\infty}^{\infty} \rmd\theta\,\frac{1}{2\pi}\int_{-\infty}^{\infty} \rmd\vartheta\,\exp(-i\theta\vartheta)\frac{1}{(2\pi)^{n-1}}[\varphi*\cdots*\varphi](\vartheta) ,
\end{align}
where there are $n-1$ convolutions of the pulse function denoted by an asterisk and $\mathcal{F}[\phi(\theta)]=\varphi(\vartheta)$ denotes the Fourier transform. By first performing the integration over $\theta$ this simplifies to
\begin{equation}
I_n(\sigma,\lambda) = \frac{1}{(2\pi)^{n-1}}[\varphi*\cdots*\varphi](0) .
\end{equation}
The two lowest order pulse integrals turn out to be independent of the pulse parameters, $I_1=1$ and 
\begin{equation}
I_2 = \int_{-\infty}^{\infty} \rmd\theta[\phi(\theta;1,\sigma,\lambda)]^2 = \frac{1}{2\pi}\int_{-\infty}^{\infty} \rmd\vartheta\,|\varphi(\vartheta;1,\sigma,\lambda)|^2 = \frac{1}{2\pi} .
\end{equation}
consistent with the requirement given by \Eqref{duration} and Parseval's theorem. From this it follows that for fixed pulse asymmetry and location parameters, the frequency power spectral density for the pulse function and its normalized auto-correlation function are the same as for a symmetric Lorentzian function,
\begin{subequations} \label{rhovarrhosigma}
\begin{align}
\Varrho_\phi(\vartheta;1,\sigma,\lambda) & = \frac{1}{I_2}\abs{\varphi(\vartheta;1,\sigma,\lambda)}^2 = \Varrho_\phi(\vartheta) ,
\\
\rho_\phi(\theta;1,\sigma,\lambda) & = \frac{1}{2\pi}\int_{-\infty}^{\infty} \rmd\vartheta\,\Varrho_\phi(\vartheta;1,\sigma,\lambda)\exp(i\theta\vartheta) = \rho_\phi(\theta) ,
\end{align}
\end{subequations}
given by \Eqsref{RvarphiL}. It is concluded that a stochastic process given by a super-position of uncorrelated Lorentzian pulses with a fixed but skewed shape has an exponential frequency power spectrum. However, it should be noted that the pulse asymmetry modifies the higher order pulse integrals $I_3$ and $I_4$, so the skewness and flatness moments given in \Eqsref{SandF-L} are only valid for symmetric Lorentzian pulses.

\begin{figure}
\includegraphics[width=8.5cm]{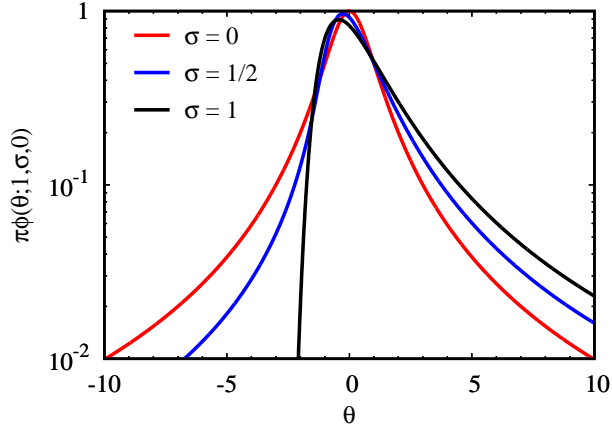}
\caption{Skewed Lorentzian pulse functions with vanishing location parameter and various values of the asymmetry parameter $\sigma$.}
\label{fig:skewlorentzian}
\end{figure}

For any function $f(\vartheta)$ with the property $f^{\dag}(\vartheta)=f(-\vartheta)$, where the dagger denotes the complex conjugate, the following relation holds,
\begin{equation}\label{ccreal}
\int_{-\infty}^{\infty} \rmd\vartheta\,f(\vartheta) = 2 \int_0^\infty \rmd\vartheta\,\mathcal{R}[f(\vartheta)] ,
\end{equation}
where $\mathcal{R}$ denotes the real part of the argument. The integrand in \Eqref{phiinvF} with $\varphi$ given by \Eqref{FsL} has this property, and thus the skewed Lorentzian pulse function can be written in integral form as
\begin{equation}
\phi(\theta;1,\sigma,\lambda) = \frac{1}{\pi}\int_0^\infty \rmd\vartheta\,\exp(-\vartheta)\cos\left( \theta\vartheta - \lambda\vartheta + \frac{2\sigma}{\pi}\,\vartheta\ln{\vartheta} \right) .
\end{equation}
While this integral cannot be calculated in closed form, it is suitable for numerical integration. The skewed Lorentzian pulse function is presented in \Figref{fig:skewlorentzian} for various values of the asymmetry parameter and vanishing location parameter, $\lambda=0$. As $\sigma$ approaches unity, the pulse function is strongly asymmetric with a fast rise and a slow decay. In the case $\sigma=1$ and $\lambda=0$ this is known as the Landau distribution. The asymptotic tail for the skewed Lorentzian pulse is for $-1<\sigma\leq1$ a power law,
\begin{equation}
\lim_{\theta\rightarrow\infty} \frac{\pi}{2}\,\theta^2\phi(\theta;1,\sigma,\lambda) = 1 .
\end{equation}
By the reflection property, it similarly follows that for $-1\leq\sigma<1$,
\begin{equation}
\lim_{\theta\rightarrow\infty} \frac{\pi}{2}\,\theta^2\phi(-\theta;1,\sigma,\lambda) = 1 .
\end{equation}
When $\sigma=1$ the left tail of the function is not asymptotically a power law and when $\sigma=-1$ the right tail of the function is not asymptotically a power law. However, in all cases there is an exponential frequency power spectral density. The role of power law tails in the pulse function for the frequency spectrum will be further discussed below.

The stochastic process defined by \Eqref{shotnoise} can be generalized to take into account a random distribution of the pulse asymmetry and location parameters for the skewed Lorentzian pulse shapes. It is natural to define the peak value of the pulse function at $\theta=0$ for all values of $\sigma$. However, since the mode of a stable distribution is not analytically expressible, a more general relationship in the form of a joint distribution is introduced. Moreover, a joint distribution $P_{\sigma\lambda A}(\sigma,\lambda,A)$ between the pulse amplitudes and the asymmetry and location parameters is considered. The mean value of the random variable $\Phi_K(t)$ in the case of exactly $K$ pulses is then given by
\begin{multline} \label{avephiKsigma}
\langle{\Phi_K}\rangle = \int_{-\infty}^\infty \rmd A_1 \int_{-\infty}^{\infty} \rmd\lambda_1 \int_{-1}^{1}\rmd\sigma_1 P_{\sigma\lambda A}(\sigma_1,\lambda_1,A_1) \int_0^T \frac{\rmd t_1}{T}
\\
\cdots \int_{-\infty}^\infty \rmd A_K \int_{-\infty}^{\infty} \rmd\lambda_K \int_{-1}^{1}\rmd\sigma_K P_{\sigma\lambda A}(\sigma_K,\lambda_K,A_K) \int_0^T \frac{\rmd t_K}{T} \sum_{k=1}^K A_k\phi\left( \frac{t-t_k}{\taud} ; 1,\sigma_k,\lambda_k \right) .
\end{multline}
Neglecting end effects by extending the integration limits for the pulse arrival times to infinity and averaging over the number of pulses, it follows that the mean value is the same as for symmetric Lorentzian pulses, $\langle{\Phi}\rangle=\gamma\Aave$. Similarly, with reference to \Eqref{FsL} it follows that the variance for the stationary process is independent of the distribution of the location and asymmetry parameters. It follows that the mean and the variance in the case of randomly skewed Lorentzian pulses are the same as for symmetric Lorentzian pulses.

The auto-correlation function in the case of a random distribution of the pulse asymmetry and location parameters can be calculated by additionally averaging over these random variables. This is determined by the $K$ terms in the double sum of $\Phi_K(t)\Phi_K(t+r)$ with equal pulse indices, given by\cite{gt-l,garcia-psd}
\begin{equation}
\sum_{k=1}^K \int_{-\infty}^{\infty} \rmd A_k \int_{-\infty}^{\infty} \rmd\lambda_k \int_{-1}^{1} \rmd\sigma_k\,A_k^2 P_{\sigma\lambda A}(\sigma_k,\lambda_k,A_k) \int_0^T\frac{\rmd t_k}{T}\phi\left( \frac{t-t_k}{\taud} ; 1,\sigma_k,\lambda_k \right)\phi\left( \frac{t-t_k+r}{\taud} ; 1,\sigma_k,\lambda_k \right) .
\end{equation}
Using the relations given by \Eqref{rhovarrhosigma}, it follows that the auto-correlation function and the frequency power spectral density for a random distribution of the pulse location and asymmetry parameters are the same as in the case with a symmetric Lorentzian pulse shape, given by \Eqref{RPhiL}. Thus, a correlation between the pulse amplitude and skewness does not influence the frequency power spectrum.

Finally, as an alternative smooth pulse function which also results in an exponential frequency power spectrum, consider the sech function,
\begin{equation}
\phi(\theta) = \frac{1}{\pi}\,\sech{(\theta)} ,
\end{equation}
which has the Fourier transform
\begin{equation}
\varphi(\vartheta) = \sech\left( \frac{\vartheta}{2} \right) .
\end{equation}
Both the pulse function and its Fourier transform have exponential tails. The integral of the $n$'th power of this pulse function is given by
\begin{equation}
I_n = \frac{\Gamma(n/2)}{\pi^{n-1/2}\Gamma(n/2+1/2)} ,
\end{equation}
giving the mean $\Phiave=\gamma\Aave$ and variance $\Phirms^2=2\gamma\langle{A^2}\rangle/\pi^2$. The normalized auto-correlation function is $\rho_\phi(\theta)=\theta\csch(\theta)$, for which the Fourier transform gives $\Varrho_\phi(\vartheta)=\pi^2/[1+\cosh(\pi\vartheta)]$.
Thus, in the case of a super-position of uncorrelated sech pulses with constant duration, the normalized auto-correlation function and power spectral density are given by
\begin{subequations}
\begin{align}
R_{\wt{\Phi}}(r) & = \frac{r}{\taud}\,\csch\left( \frac{r}{\taud} \right) ,
\\
\frac{1}{\taud}\,\Omega_{\wt{\Phi}}(\omega) & = \frac{\pi^2}{1+\cosh(\pi\taud\omega)} .
\end{align}
\end{subequations}
The auto-correlation function has an exponential tail for large time lags and the frequency power spectral density has the asymptotic limits
\begin{subequations}
\begin{gather}
\lim_{\taud\abs{\omega}\rightarrow0} \frac{2}{\pi^2}\frac{\Omega_{\wt{\Phi}}(\omega)}{\taud} = 1 ,
\\
\lim_{\taud\abs{\omega}\rightarrow\infty} \frac{\exp(\pi\taud\abs{\omega})}{2\pi^2} \frac{\Omega_{\wt{\Phi}}(\omega)}{\taud} = 1 ,
\end{gather}
\end{subequations}
decreasing exponentially for high frequencies, qualitatively similar to the case of Lorentzian pulses. However, the spectrum for sech pulses is flat for low frequencies. It is concluded that the algebraic tails of the Lorentzian pulse function leads to a high power spectral density at low frequencies. The exponential spectrum for high frequencies reflects the curvature of the smooth sech and Lorentzian pulses and is not due to algebraic tails in the pulse function.

Intermittent fluctuations in physical systems have been investigated by a stochastic model that describe these as a super-position of uncorrelated pulses. It is demonstrated that for skewed Lorentzian pulses with constant duration, the spectrum is unaffected by the asymmetry parameter even in the case when this has a random distribution and is correlated with the pulse amplitudes. Such correlations are expected due to amplitude dependent steepening of smooth pulses resulting from non-linear advection in continuum systems. The results presented here provide new insight to the ubiquitous appearance of exponential spectra in fluctuating fluids and plasmas, in particular demonstrating that the power spectral density is not influenced by the steepening process in the case of Lorentzian pulses. Previously, it has been shown that a broad distribution of pulse duration times leads to frequency spectra with power law scaling regimes.\cite{gt-l}


This work was supported with financial subvention from the Research Council of Norway under grant 240510/F20. The authors acknowledge the generous hospitality of the MIT Plasma Science and Fusion Center where parts of this work was conducted.

\end{document}